\documentclass[journal,letterpaper,twoside,twocolumn]{IEEEtran}
\usepackage{amsmath,amssymb,amscd,latexsym,dsfont,mathtools}
\usepackage{float}
\usepackage{color}
\usepackage{graphicx}
\usepackage{comment}
\usepackage{subfigure}
\usepackage{enumerate}
\usepackage{stfloats}
\usepackage{psfrag}
\usepackage{setspace}
\usepackage{cite}
\usepackage{balance}
\usepackage{multirow}
\usepackage{rotating}
\usepackage[para]{threeparttable}
\usepackage{pstricks}
\usepackage{pstricks-add}
\usepackage{pst-plot}
\usepackage{pst-sigsys}
\usepackage{colortbl}
\usepackage{slashbox}
\usepackage{pbox}
\usepackage{lipsum}

\IEEEoverridecommandlockouts

\renewcommand{\markboth}[1]{\renewcommand{\leftmark}{#1}\renewcommand{\rightmark}{#1}}
\newcommand{\argmin}[2]{\mathop{\mathrm{argmin}}_{#1} {#2}}
\newcommand{\argmax}[2]{\mathop{\mathrm{argmax}}_{#1} {#2}}
\newcommand{\qfunc}[1]{\mathrm{Q}\left( #1 \right)}
\newcommand{\trans}{\mathsf{T}}

\newtheorem{theorem}{Theorem}

\newtheorem{remark}{Remark}

\newtheorem{corollary}{Corollary}
\newtheorem{lemma}{Lemma}

\newcommand{\bbb}{\boldsymbol{b}}
\newcommand{\qqq}{\boldsymbol{q}}
\newcommand{\ccc}{\boldsymbol{C}}
\newcommand{\xxx}{\boldsymbol{x}}

\newcommand{\llll}{\boldsymbol{L}}
\newcommand{\yyy}{\boldsymbol{Y}}

\newcommand{\eee}{\boldsymbol{e}}
\newcommand{\HHH}{\boldsymbol{H}}
\newcommand{\setX}{\mathcal{X}}
\newcommand{\setS}{\mathcal{S}}
\newcommand{\setB}{\mathcal{B}}
\newcommand{\setE}{\mathcal{E}}
\newcommand{\setK}{\mathcal{K}}
\newcommand{\PhiB}{\Phi_{\setB}}
\newcommand{\PhiX}{\Phi_\setX}
\newcommand{\PhiS}{\Phi_\setS}
\newcommand{\PEP}{\mathrm{PEP}}
\renewcommand{\S}{\setX}
\newcommand{\B}{\setB}

\newcommand{\Loss}{\mathsf{L}}
\newcommand{\CLoss}{\mathsf{L}}

\newcommand{\GC}[1]{GL\ensuremath{_#1}}

\newcommand{\eqsref}[2]{(\ref{#1})--(\ref{#2})}
\newcommand{\figref}[1]{Fig.~\ref{#1}}

\newcommand{\tabref}[1]{Table~\ref{#1}}
\newcommand{\tabsref}[2]{Tables~\ref{#1} and \ref{#2}}
\newcommand{\theref}[1]{Theorem~\ref{#1}}
\newcommand{\thesref}[2]{Theorems~\ref{#1} and \ref{#2}}
\newcommand{\secref}[1]{Sec.~\ref{#1}}

\title{On the Asymptotic Performance of Bit-Wise Decoders for Coded Modulation}
\author{%
\IEEEauthorblockN{Mikhail Ivanov, Alex Alvarado{\IEEEauthorrefmark{4}}, Fredrik Br\"{a}nnstr\"{o}m, Erik Agrell\\}
\IEEEauthorblockA{Department of Signals and  Systems, Chalmers  University of Technology, Gothenburg, Sweden\\} \IEEEauthorblockA{ \IEEEauthorrefmark{4}Department of Engineering, University of Cambridge, UK\\ \emph{\{mikhail.ivanov,fredrik.brannstrom,agrell\}@chalmers.se, alex.alvarado@ieee.org}
}

\thanks{Research supported by the Swedish Research Council, Sweden (under grant \#621-2011-5950), by the Ericsson's Research Foundation, Sweden (under grant \#556016-0680), and by the European Community's Seventh Framework Programme (FP7/2007-2013) under grant agreement No. 271986. The calculations were performed on resources provided by the Swedish National Infrastructure for Computing (SNIC) at C3SE.}
}

\begin{document}
\maketitle

\begin{abstract}
Two decoder structures for coded modulation over the Gaussian and flat fading channels are studied: the maximum likelihood symbol-wise decoder, and the (suboptimal) bit-wise decoder based on the bit-interleaved coded modulation paradigm. We consider a 16-ary quadrature amplitude constellation labeled by a Gray labeling. It is shown that the asymptotic loss in terms of pairwise error probability, for any two codewords caused by the bit-wise decoder, is bounded by 1.25~dB. The analysis also shows that for the Gaussian channel the asymptotic loss is zero for a wide range of linear codes, including all rate-1/2 convolutional codes.
\end{abstract}

\begin{IEEEkeywords}
Additive white Gaussian noise, flat fading channel, Gray code, pairwise error probability, coded modulation, bit-interleaved coded modulation, logarithmic likelihood ratio, pulse-amplitude modulation.
\end{IEEEkeywords}

\section{Introduction and Motivation}\label{sec:intro}

Coded modulation (CM) is a concatenation of multilevel modulation and a channel code. One popular coded modulation scheme was proposed and analyzed~in \cite{Ungerboeck76, Unger82jan}, where convolutional codes (CCs) were used. Due to the trellis structure of the resulting codes, such systems are called trellis-coded modulation (TCM). The TCM decoder finds the codewords at minimum Euclidean distance by exploiting the trellis structure of the code, e.g., by using a symbol-by-symbol Viterbi algorithm. Around the same time, multilevel coding (MLC) was presented in~\cite{Imai77}, where the main idea was to use different binary codes for different bit positions of the constellation points and multiple decoders at the receiver.

Bit-interleaved coded modulation (BICM) is another approach for CM. BICM was initially proposed in~\cite{Zehavi92may} and later studied in~\cite{Fabregas08_Book,Caire98}. In BICM, the encoder and the modulator are separated by a bit-level interleaver. At the receiver side, a suboptimal bit-wise decoder is used, which operates on the L-values provided by the demapper.

It has recently been shown in~\cite{Stierstorfer10} (see also \cite{Alvarado10d}) that removing the interleaver improves the performance of BICM over the additive white Gaussian noise (AWGN) channel. Somewhat surprisingly, the results in \cite{Alvarado10d} show that for CCs, the performance of a bit-wise decoder for an optimized BICM system without an interleaver is asymptotically equivalent to the performance of an optimized TCM system. As~\cite{Alvarado13} reveals, these two optimized systems use the same transmitters, i.e., the symbol sequences going into the channel are the same, even though they use different convolutional encoders and binary labelings.

In this paper, we generalize the results in \cite{Stierstorfer10,Alvarado10d} by studying the asymptotic difference between symbol-wise and bit-wise decoders for CM systems with arbitrary binary linear encodes. We consider $16$-ary quadrature amplitude modulation (QAM) with a Gray labeling over the AWGN, as well as over flat fading channels. The main result of the paper consists in showing that for any two codewords, the pairwise error probability (PEP) loss caused by the bit-wise decoder is bounded by $1.25$~dB. We also prove that for a wide range of linear codes, the asymptotic loss caused by the bit-wise decoder is zero over the AWGN channel.

\section{System Model}\label{sec:syst_mod}

\subsection{Coded Modulation Encoder}

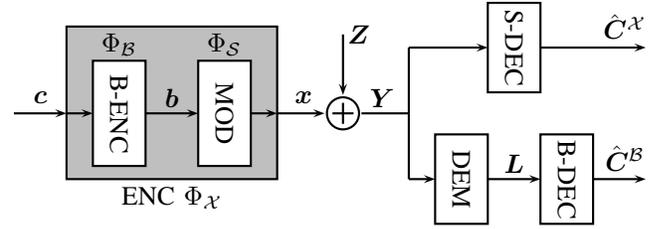
\begin{figure}[tp]
\begin{center}
	\psset{unit=.35cm}
	\begin{pspicture}(-1,-1.5)(23,7.5)
        \pspolygon[fillstyle = solid, fillcolor = lightgray, linearc=0](1, 0.5)(1, 6.3)(9, 6.3)(9, 0.5) 
        \pspolygon[fillstyle = solid, fillcolor = white, linearc=0](2, 1)(2, 5)(4, 5)(4, 1)  
        \pspolygon[fillstyle = solid, fillcolor = white, linearc=0](6, 1)(6, 5)(8, 5)(8, 1) 
        \psline{->}(-1, 3)(1, 3)
        \psline{->}(1, 3)(2, 3)
        \psline{->}(4, 3)(6, 3)
        \psline{->}(8, 3)(9, 3)
        \psline{->}(9, 3)(10.8, 3)
	\pscircle(11.5,3){0.65}
	\psline{-}(11.1, 3)(11.9, 3)
	\psline{-}(11.5, 2.6)(11.5, 3.4)
        \psline{-}(12.2, 3)(14, 3)
        \psline{->}(11.5, 6)(11.5, 3.7)
        \pspolygon[linearc=0](17, 3.8)(17, 7.2)(19, 7.2)(19, 3.8)
        \pspolygon[linearc=0](15, -1.2)(15, 2.2)(17, 2.2)(17, -1.2)
        \pspolygon[linearc=0](19, -1.2)(19, 2.2)(21, 2.2)(21, -1.2)
        \psline{-}(14, 3)(14, 5.5)
        \psline{-}(14, 3)(14, 0.5)
        \psline{->}(14, 5.5)(17, 5.5)
        \psline{->}(14, 0.5)(15, 0.5)
        \psline{->}(19, 5.5)(23, 5.5)
        \psline{->}(17, 0.5)(19, 0.5)
        \psline{->}(21, 0.5)(23, 0.5)
        \rput{-90}(3, 3){B-ENC}
        \rput(3, 5.6){$\PhiB$}
	\rput{-90}(7, 3){MOD}
        \rput(7, 5.6){$\PhiS$}
        \rput(5, -0.2){ENC $\PhiX$}
        \rput{-90}(18, 5.5){S-DEC}
        \rput{-90}(16, 0.5){DEM}
        \rput{-90}(20, 0.5){B-DEC}
        \rput(0, 3.6){$\boldsymbol{c}$}
        \rput(5, 3.6){$\bbb$}
        \rput(10, 3.6){$\xxx$}
        \rput(13, 3.6){$\yyy$}
        \rput(22.2, 6.3){$\hat{\ccc}{}^\S$}
        \rput(22.2, 1.3){$\hat{\ccc}{}^\B$}
        \rput(18, 1.1){$\llll$}
        \rput(12.1, 6){$\boldsymbol{Z}$}
	\end{pspicture}
\end{center}
\caption{Block diagram of the analyzed CM system. The CM encoder $\PhiX$ is used at the transmitter. At the receiver,  two decoding algorithms are considered: the ML symbol-wise decoder S-DEC or a suboptimal bit-wise decoder B-DEC.}
\label{fig:block_scheme}
\end{figure}

Throughout the paper, boldface letters denote vectors or matrices and capital letters denote random variables. The block diagram of the analyzed system is shown in~\figref{fig:block_scheme}. A CM encoder (ENC) carries out a one-to-one mapping from an information vector of $K$ bits $\boldsymbol{c} = [c[1],\dots, c[K]]\in\{0,1\}^K$ to a vector of $N$ symbols $\xxx = [x[1],\dots, x[N]]$. Each symbol is drawn from a discrete constellation $\setS=\{s_1,\dots, s_M\}$, i.e., $x[k] \in \setS$ and $k = 1,\dots,N$, where $M = 2^m$ and $m$ is a positive integer. All vectors $\xxx$ form a \emph{CM code} $\setX \subset \setS^{N}$, where $|\setX| = 2^K$ is the number of possible information vectors. The CM encoder is defined as the function $\PhiX: \{0, 1\}^K \rightarrow \setX$ with the corresponding inverse function $\PhiX^{-1}: \setX \rightarrow \{0, 1\}^K$. Assuming all information vectors to be equally likely, the average energy per symbol can be expressed as $E_s = N^{-1}2^{-K}\sum_{\xxx \in \setX}{\|\xxx\|^2}$ and the average energy per bit $E_b = K^{-1}NE_s$.

As all symbols $s_i$ can be uniquely identified by length-$m$ binary labels, any CM encoder described above can be represented as a concatenation of two blocks, as shown in~\figref{fig:block_scheme}. The modulator (MOD) carries out a one-to-one mapping from $m$ bits to one of the $M$ constellation points. The modulator is defined as the function $\PhiS: \{0, 1\}^m \rightarrow \setS$ with the corresponding inverse function $\PhiS^{-1}: \setS\rightarrow \{0, 1\}^m $. We represent a binary labeling by a vector $\qqq = [q_1,\dots, q_M]$, where $q_i$ is the integer representation of the $m$ bits mapped to the symbol $s_i$, with the most significant bit to the left.

A binary encoder (B-ENC) provides  the modulator with bits to produce a vector of symbols $\xxx$. The B-ENC maps $K$ incoming bits $\boldsymbol{c}$ into $mN$ coded bits $\bbb = [\bbb[1],\dots, \bbb[N]]$, where $\bbb[k] = [b_1[k],\dots,b_m[k]] = \PhiS^{-1}(x[k]) \in \{0,1\}^m$ and $k=1,\dots,N$. All vectors $\bbb$ form a \emph{binary code} $\setB \subset \{0,1\}^{mN}$, where $|\setB| = |\setX| = 2^K$. The B-ENC is defined as the function $\PhiB: \{0, 1\}^K \rightarrow \setB$ with the corresponding inverse function $\PhiB^{-1}: \setB \rightarrow \{0, 1\}^K$. Throughout the paper, we assume $\setB$ to be a binary \emph{linear} code.

The described CM encoder in~\figref{fig:block_scheme} generalizes the proposed coding schemes in~\cite{Unger82jan, Imai77, Zehavi92may}. Indeed, it corresponds to TCM if the B-ENC is a terminated convolutional encoder. If the B-ENC is a bank of $m$ parallel  encodes, the described encoder represents an MLC encoder. Finally, it corresponds to BICM if the B-ENC includes an interleaver. For rate-$1/2$ CCs, the considered setup is similar to the one considered in~\cite{Alvarado10d} except for the fact that no random scrambling of the coded bits (see ~\cite[Sec.~II]{Alvarado10d} for more details) is used in this paper.

When using binary phase-shift keying, the function of the modulator is trivial, and analyzing the CM code $\setX$ is equivalent to analyzing a corresponding binary code $\setB$. This, however, is not the case when multilevel modulation is used.

In this paper, we study a $16$-QAM constellation formed as a direct product of two $4$-ary pulse amplitude modulation (PAM) constellations. The labeling of the $16$-QAM constellation is also obtained as a direct product of two Gray-labeled $4$-PAM constellations. This configuration is relevant in practice, as it allows to decouple the two-dimensional detection into detection of each dimension separately. This is used in many wireless standards, see e.g.,~\cite[Fig.~18-10]{IEEE80211-2012}, \cite[Table~7.1.3-1]{ETSI_TS_136211-2013}, \cite[Fig.~15]{ETSI_EN_302_755_v131}.  Therefore, only the constituent $4$-PAM constellation needs to be considered. This constellation is defined as  $\setS=\{-3d, -d, d, 3d\}$, where $d$ is a normalization factor and $s_{i}<s_{i+1}$.

We consider a real discrete-time memoryless AWGN channel, i.e., given the channel input $x$, the channel output is $Y = x + Z$, where $Z$ is a zero-mean Gaussian random variable  with variance $\sigma_z^2 = N_0/2$. The conditional probability density function (PDF) of the channel output is
\begin{equation}
    p_{Y|X}(y|x) = \frac{1}{\sqrt{2\pi \sigma_z^2}}\mathrm{e}^{-\frac{(y - x)^2}{2\sigma_z^2}}.
    \label{eq:gauss}
\end{equation}
A Gaussian distribution with mean value $\mu$ and variance $\sigma^2$ is denoted by $\mathcal{N}(\mu, \sigma^2)$, i.e.,  $Y \sim \mathcal{N}(x, \sigma_z^2)$. Flat fading channels will be discussed in~\secref{sec:fading}.

It is well known that there are $4! = 24$ labelings for $4$-PAM. Due to the symmetry of the constellation and the channel, the labelings $\qqq = [q_1,q_2, q_3, q_4]$ and $\qqq' = [q_4, q_3, q_2, q_1]$ will produce equivalent CM codes $\setX$ and $\setX'$ for any binary code $\setB$, i.e., if a codeword $\xxx$ belongs to the code $\setX$, then  $-\xxx$ belongs to the code $\setX'$. The number of labelings is therefore reduced to 12. Four of them are Gray labelings, which are listed in~\tabref{tab:Grays}. In this paper, only Gray labelings are considered.

\begin{table}
    \centering
    \caption{Gray labelings for $4$-PAM}
    \begin{tabular}{|c|c|}
        \hline
        Labeling & $\qqq$\\
        \hline
        \GC1& $[0, 1, 3, 2]$\\
        \hline
        \GC2& $[0, 2, 3, 1]$\\
        \hline
        \GC3& $[1, 0, 2, 3]$\\
        \hline
        \GC4& $[2, 0, 1, 3]$\\
        \hline
    \end{tabular}
    \label{tab:Grays}
\end{table}

The most popular Gray labeling is \GC1, often referred to as the binary reflected Gray code (BRGC)~\cite{Gray 53, Agrell04dec, Agrell07}. All these labelings give the same uncoded bit error rate and BICM generalized mutual information~\cite{Martinez09} for the AWGN channel, thus, they are usually said to be equivalent~\cite{Agrell04dec}. However, in this paper, we consider them separately, as all these labelings produce different CM codes when used with a given binary code $\setB$.

In this paper, we study two different decoders for the CM encoder in~\figref{fig:block_scheme}, which we describe below.

\subsection{Symbol-Wise Decoder}

The symbol-wise decoder (S-DEC) shown in~\figref{fig:block_scheme} performs maximum likelihood (ML) decoding by computing
\begin{equation}\label{eq:S-decoding_rule}
    \hat{\ccc}{}^\S = \Phi_{\setX}^{-1}\left(\argmin{\xxx \in \setX}{\left\{D^\S(\xxx)\right\}} \right),
\end{equation}
where $D^\S(\xxx) = \sum_{k = 1}^{N}{\left(Y[k]-x[k]\right)^2}$. In other words, the S-DEC searches for the closest codeword to the observation $\yyy = [Y[1],\dots, Y[N]]$. Assuming the codeword $\xxx \in \setX$ is transmitted, an error occurs if there is a codeword $\hat{\xxx} = [\hat{x}[1],\dots,\hat{x}[N]] \in \setX$, such that $D^\S(\xxx)> D^\S(\hat{\xxx})$. The probability of such an event is called the PEP and can be calculated as
\begin{equation}\label{eq:PEP_SDEC}
    \PEP^\S(\xxx, \hat{\xxx}) = \Pr\{\Delta^\S(\xxx, \hat{\xxx}) < 0\},
\end{equation}
where $\Pr\{\cdot\}$ stands for probability and $\Delta^\S(\xxx, \hat{\xxx}) \triangleq D^\S(\hat{\xxx})-D^\S(\xxx)$. For future use, we express $\Delta^\S(\xxx, \hat{\xxx})$ as
\begin{align}
\Delta^\S(\xxx, \hat{\xxx}) = \sum_{k = 1}^{N}\Lambda^\S(x[k], \hat{x}[k]),
\end{align}
where
\begin{equation}
    \Lambda^\S(x[k], \hat{x}[k]) =  2(x[k] -\hat{x}[k])Y[k] + \hat{x}^2[k]-x^2[k]         \label{eq:cm_lambda}
\end{equation}
is called a symbol metric difference (SMD).

\subsection{Bit-Wise Decoder}
The bit-wise decoder (B-DEC) shown in~\figref{fig:block_scheme} operates on the bit reliability metrics provided by a demapper (DEM). The demapper acts independently of the B-DEC and calculates a vector $\llll = [\llll[1],\dots, \llll[N]]$, where $\llll[k] = [L_1[k],\dots, L_m[k]]$ are the logarithmic-likelihood ratios (L-values). We use the so-called max-log approximation~\cite[eq.~(3.2)]{Zehavi92may}, \cite[eq.~(2.15)]{Fabregas08_Book}, \cite[eq.~(12)]{Martinez09} for the calculation of the L-values, i.e.,
\begin{equation}
L_j[k] = \frac{1}{2\sigma_z^2}\left[\min_{s \in \setS_{j, 0}}{(Y[k]-s)^2} - \min_{s \in \setS_{j, 1}}{(Y[k]-s)^2}\right]
    \label{eq:max_log_rrl}
\end{equation}
with $j=1,\dots,m$, where $\setS_{j,u} \subset \setS$ is the subset of constellation points whose labels have the value $u \in \{0, 1\}$ in the $j$th bit position.

The calculated L-values are passed to the B-DEC, which uses the decoding rule \cite[Sec.~2.2]{Fabregas08_Book}, \cite[eq.~(13)]{Martinez09}	
\begin{equation}
    \hat{\ccc}{}^\B = \Phi_{\setB}^{-1}\left(\argmax{\bbb \in \setB}{\left\{D^\B(\bbb)\right\}} \right),
    \label{eq:dec_llrs}
\end{equation}
where $D^\B(\bbb) = (2\bbb-1)\llll^\trans = \sum_{k = 1}^{N}(2\bbb[k]-1) \llll^\trans[k]$ and $(\cdot)^\trans$ denotes transposition.

The PEP for the B-DEC is given by
\begin{equation} \label{eq:PEP_BDEC}
    \PEP^\B(\bbb, \hat{\bbb}) = \Pr\{\Delta^\B(\bbb, \hat{\bbb})<0\},
\end{equation}
where $\Delta^\B(\bbb, \hat{\bbb}) \triangleq D^\B(\bbb)-D^\B(\hat{\bbb})$ is the difference between the metrics for the transmitted codeword $\bbb$ and the competing codeword $\hat{\bbb}\in \setB$. Since the mapping between $\bbb$ and $\xxx$ is one-to-one, with a slight abuse of notation, $\Delta^\B(\bbb, \hat{\bbb})$ can be written as a function of codewords $\xxx$ and $\hat{\xxx}$ instead, i.e., 
\begin{equation} \label{eq:dist_bdec}
\Delta^\B(\xxx, \hat{\xxx}) = \sum_{k = 1}^{N}{\Lambda^\B(x[k], \hat{x}[k])},
\end{equation}
where the SMD in this case is
\begin{equation}
    \Lambda^\B(x[k], \hat{x}[k]) = 2(\PhiS^{-1}(x[k]) - \PhiS^{-1}(\hat{x}[k]))\llll[k]^\trans.
    \label{eq:delta_llrs}
\end{equation}

The bit-wise decoder described above corresponds to the standard (noniterative) BICM decoder. We refrain from using this name, as the interleaver might or might not be included in the transmitter. Moreover, if there is an interleaver, we assume it to be part of the B-ENC.

\section{Symbol vs. Bit Decoder}\label{sec:comparison}
\subsection{Distribution of the SMDs}

To compare the PEP for the S-DEC in~\eqref{eq:PEP_SDEC} and the B-DEC in~\eqref{eq:PEP_BDEC}, we analyze the distributions of the SMDs in~\eqref{eq:cm_lambda} and~\eqref{eq:delta_llrs}.

\begin{lemma}\label{SMDs_SDEC.Lemma}
For $4$-PAM with any labeling, the SMDs in \eqref{eq:cm_lambda} divided by $4d$ are distributed as
    \begin{equation}
        (4d)^{-1}\Lambda^{\S}(x[k], \hat{x}[k]) \sim \mathcal{N}(\mu d, \sigma^2 \sigma^2_z),
    \end{equation}
where $(\mu, \sigma^2)$ are shown in~\tabref{tab:S_distributions}.
\end{lemma}

\begin{table}
    \centering
    \caption{Distribution parameters $(\mu, \sigma^2)$ for the SMD~\eqref{eq:cm_lambda} of the S-DEC. Circles, stars, and diamonds show the error vector $\eee$ equal to $[0,1]$, $[1,1]$, and $[1,0]$, respectively, for \GC3.}
    \begin{tabular}{|c|c|c|c|c|}
        \hline
        $x[k]$ \,\, $\hat{x}[k]$& $s_1$ & $s_2$ & $s_3$ & $s_4$\\
        \hline
        $s_1$ & -- & $(1, 1)^{\circ}$ & $(4, 4)^{\star}$ & \cellcolor[gray]{0.9}$(9, 9)^{\diamond}$ \\
        \hline
        $s_2$ & $(1, 1)^{\circ}$ & --& $(1, 1)^{\diamond}$ & $(4, 4)^{\star}$ \\
        \hline
        $s_3$ & $(4, 4)^{\star}$ & $(1, 1)^{\diamond}$ & -- & $(1, 1)^{\circ}$ \\
        \hline
        $s_4$ &\cellcolor[gray]{0.9}$(9, 9)^{\diamond}$ & $(4, 4)^{\star}$ & $(1, 1)^{\circ}$ & -- \\
        \hline
    \end{tabular}
    \label{tab:S_distributions}
    \psline[linewidth = 0.02](-6.4, 0.74) (-6.79, 1.18)
\end{table}

\begin{IEEEproof}
Since the SMDs in~\eqref{eq:cm_lambda} are linear functions of the observation $Y[k]$, the SMDs follow a Gaussian distribution. When $x[k] = s_i$ and $\hat{x}[k] = s_j$, the mean value of the scaled SMD is $\mu = (4d^2)^{-1}(2(s_i - s_j)s_i + (s_j^2 - s_i^2)) = (4d^2)^{-1}(s_i - s_j)^2$. The variance can be calculated as $\sigma^2 = (4 d^2)^{-1}(s_i - s_j)^2$. Substituting values of $s_i$ and $s_j$ gives the parameters shown in~\tabref{tab:S_distributions}.
\end{IEEEproof}

We note that the results in~Lemma~\ref{SMDs_SDEC.Lemma} are valid for any labeling, not only Gray labelings. Scaling of the SMDs in~\eqref{eq:cm_lambda} by $4d$ changes neither the performance of the S-DEC nor the analysis. However, it simplifies the notation and makes the comparison of the S-DEC and the B-DEC clearer. For the same reasons, the SMDs in~\eqref{eq:delta_llrs} are scaled by $\sigma^2_z$ in the following lemma.

\begin{lemma}\label{SMDs_BDEC.Lemma}
For $4$-PAM with any Gray labeling, the distribution of the SMDs in~\eqref{eq:delta_llrs} scaled by $\sigma^2_z$ can be approximated as
    \begin{equation}
        \sigma^2_z\Lambda^{\B}(x[k], \hat{x}[k]) \sim \mathcal{N}(\mu d, \sigma^2 \sigma^{2}_z),
    \end{equation}
where $(\mu, \sigma^2)$ are shown in~\tabref{tab:B_distributions}.
\end{lemma}

\begin{table}
    \centering
    \caption{Distribution parameters $(\mu, \sigma^2)$ for the SMD~\eqref{eq:delta_llrs} of the B-DEC. Circles, stars, and diamonds show the error vector $\eee$ equal to $[0,1]$, $[1,1]$, and $[1,0]$, respectively, for \GC3.}
    \begin{tabular}{|c|c|c|c|c|}
        \hline
       $x[k]$ \,\,  $\hat{x}[k]$& $s_1$ & $s_2$ & $s_3$ & $s_4$\\
        \hline
        $s_1$ & -- & $(1, 1)^{\circ}$ & $(4, 4)^{\star}$ &\cellcolor[gray]{0.9}$(3, 1)^{\diamond}$ \\
        \hline
        $s_2$ & $(1, 1)^{\circ}$ & --& $(1, 1)^{\diamond}$ & $(4, 4)^{\star}$ \\
        \hline
        $s_3$ & $(4, 4)^{\star}$ & $(1, 1)^{\diamond}$ & -- & $(1, 1)^{\circ}$ \\
        \hline
        $s_4$ &\cellcolor[gray]{0.9}$(3, 1)^{\diamond}$ & $(4, 4)^{\star}$ & $(1, 1)^{\circ}$ & -- \\
        \hline
    \end{tabular}
    \label{tab:B_distributions}
    \psline[linewidth = 0.02](-6.4, 0.74) (-6.79, 1.18)
\end{table}

\begin{IEEEproof}
Since the L-value in~\eqref{eq:max_log_rrl} is a piece-wise linear function of the observation, the distribution of the L-value is a superposition of piece-wise Gaussian distributions, with mean and variance defined by the linear pieces and the transmitted symbol. In~\cite[Sec.~5]{Benjillali07}, \cite[Sec.~III-C]{Alvarado07d}, it has been shown that at high signal-to-noise ratios (SNR), measured as $E_s/N_0$ or $E_b/N_0$, the so-called zero-crossing (ZcMod) approximation of such a PDF gives good results in terms of coded bit-error rate (BER) and mutual information. The results shown in~\tabref{tab:B_distributions} are obtained from \cite[Table~II]{Alvarado10d} by scaling the SMDs by $\sigma^2_z$. The distributions are independent of a particular Gray labeling and depend only on the compared symbols. The tightness of the ZcMod approximation will be discussed in~\secref{sec:ZcMod}.
\end{IEEEproof}

Comparing~\tabsref{tab:S_distributions}{tab:B_distributions}, we note that the tables are identical, except for the corner entries in gray. We will use this simple observation in the following section to bound the loss incurred by the B-DEC when compared to the S-DEC.

\subsection{Pairwise Error Probability Analysis}\label{sec:pep}

In this section, we study the asymptotic performance of the S-DEC and the B-DEC. Throughout the section, we use \GC3 for illustration, i.e., symbols $s_k$, $k = 1,\dots, 4$ are labeled with $[0,1]$, $[0,0]$, $[1,0]$, and $[1,1]$, respectively.  All discussions and derivations below apply directly to \GC1, and also to \GC2 and \GC4 if the labels $[1,0]$ and ${[0,1]}$ are swapped.

Examining~\tabsref{tab:S_distributions}{tab:B_distributions}, we see that, in many cases, the distribution of the SMDs depends on the binary vector $\eee \triangleq \Phi^{-1}_{\mathcal{S}}(x[k]) \oplus \Phi^{-1}_{\mathcal{S}}(\hat{x}[k])\,\, \in \{0,1\}^2$, where $\oplus$ denotes modulo-2 addition. 
When $\eee = [0, 0]$, the distributions are not defined (main diagonal of the tables). For  $\eee = [1, 1]$, the distribution parameters are $(4, 4)$ (marked with stars in the tables) and for $\eee = [0, 1]$, the distribution parameters are $(1, 1)$ (marked with circles). However, the distribution parameters for $\eee = [1, 0]$ are different (marked with diamonds in the tables). When the compared symbols are $s_2$ and $s_3$, the distribution parameters are $(1, 1)$, whereas the distribution parameters are $(9, 9)$ and $(3, 1)$ for the S-DEC and the B-DEC, respectively, when the compared symbols are $s_1$ and $s_4$ (gray entries of the tables). We use $(\mu_{[0,1]}, \sigma_{[0,1]}^2)$ for entries marked with circles, $(\mu_{[1,1]}, \sigma_{[1,1]}^2)$ for entries marked with stars, $(\mu_{[1,0]}, \sigma_{[1,0]}^2)$ for white entries marked with diamonds, and 
$(\mu_{\setX}, \sigma_{\setX}^2)$ and $(\mu_{\setB}, \sigma_{\setB}^2)$ for gray entries marked with diamonds for the S-DEC and the B-DEC, respectively.

We define the set of possible non-zero vectors $\eee$ as $\setE = \{[0, 1], [1,0], [1, 1]\}$. For two codewords $\xxx$ and $\hat{\xxx}$ and for $\eee \in \setE$, we define $w_{\eee}(\xxx, \hat{\xxx})$ as 
\begin{equation}
    w_{\eee}(\xxx, \hat{\xxx}) = \sum_{k = 1}^{N}\mathrm{I}\left\{\Phi^{-1}_{\mathcal{S}}(x[k]) \oplus \Phi^{-1}_{\mathcal{S}}(\hat{x}[k]) = \eee\right\},
\end{equation}
where $\mathrm{I}\{\cdot\}$ is the indicator function. In other words, $w_{\eee}(\xxx, \hat{\xxx})$ is the number of pairs $(x[k], \hat{x}[k])$ in $\xxx$ and $\hat{\xxx}$ such that $\Phi^{-1}_{\mathcal{S}}(x[k]) \oplus \Phi^{-1}_{\mathcal{S}}(\hat{x}[k]) = \eee$. In addition, we define $w_{c}(\xxx, \hat{\xxx})$ as the number of pairs $(x[k], \hat{x}[k])$ in $\xxx$ and $\hat{\xxx}$ such that $(x[k], \hat{x}[k]) = (s_1, s_4)$ or $(x[k], \hat{x}[k]) = (s_4, s_1)$, i.e., $w_{c}(\xxx, \hat{\xxx})$ is the number of \emph{corner} entries (gray entries in \tabsref{tab:S_distributions}{tab:B_distributions}). 
Clearly, $w_{[1,0]}(\xxx, \hat{\xxx}) \ge w_{c}(\xxx, \hat{\xxx})$, as the former includes pairs of symbols counted in the latter. To simplify the notation, the arguments of $w_{\eee}(\xxx, \hat{\xxx})$ and $w_{c}(\xxx, \hat{\xxx})$ are omitted when the arguments are clearly stated in the text.

From Lemmas~1 and~2, it follows that the SMDs are independent Gaussian random variables. Using the introduced notation, the PEP for the S-DEC and the B-DEC in~\eqref{eq:PEP_SDEC} and~\eqref{eq:PEP_BDEC} can therefore be expressed as
\begin{equation} \label{eq:PEP}
    \PEP(\xxx, \hat{\xxx}) = \qfunc{a(\xxx, \hat{\xxx})\frac{d}{\sigma_z}},
\end{equation}
where $\qfunc{\cdot}$ is the Gaussian Q-function and the normalized distance $a(\xxx, \hat{\xxx})$ is either
\begin{align}\label{A.PEP}
    a^{\setX}(\xxx, \hat{\xxx}) & = \frac{ w_{c}(\mu_{\setX} - \mu_{[1,0]}) + \sum_{\eee \in \setE}{w_{\eee}\mu_{\eee}}   } {\sqrt{w_{c}(\sigma^2_{\setX} - \sigma^2_{[1,0]})+   \sum_{\eee \in \setE}{w_{\eee}\sigma^2_{\eee}}  }}
\end{align}
for the S-DEC or 
\begin{align}\label{A.PEP2}
    a^{\setB}(\xxx, \hat{\xxx}) & = \frac{ w_{c}(\mu_{\setB} - \mu_{[1,0]}) + \sum_{\eee \in \setE}{w_{\eee}\mu_{\eee}}   } {\sqrt{w_{c}(\sigma^2_{\setB} - \sigma^2_{[1,0]})+   \sum_{\eee \in \setE}{w_{\eee}\sigma^2_{\eee}}  }}
\end{align}
for the B-DEC.


\begin{figure}
    \newcommand{\scale}{0.85}
    \psfrag{xlabel}[cc][cB][\scale]{$d/\sigma_z$~[dB]}%
    \psfrag{ylabel}[cc][ct][\scale]{PEP}%
    \psfrag{S-DEC111}[c][c][\scale]{\hspace{-2mm}S-DEC}
    \psfrag{B-DEC111}[c][c][\scale]{\hspace{-2mm}B-DEC}%
    \centering
    \includegraphics[width=\columnwidth]{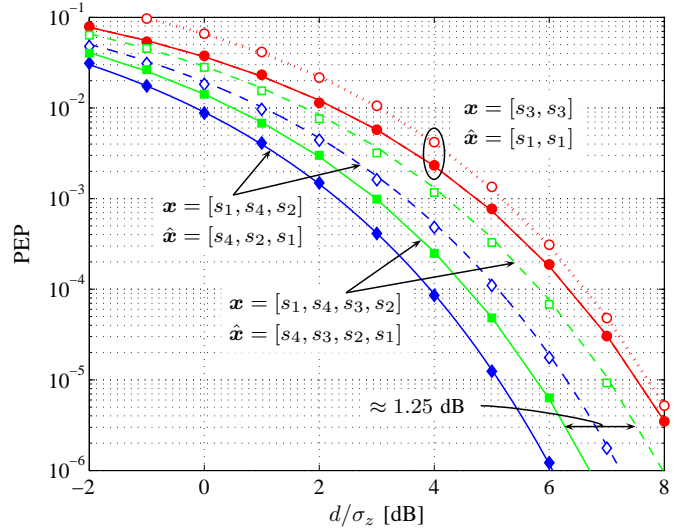}
    \vspace{-0.42cm}
	\begin{pspicture}(0,0)(0,0)
	\psset{linewidth=.02cm}   
    \rput[0.2](-1.5, 4.65){\psframebox*{\footnotesize $\xxx = [s_1,s_4,s_2]$}}
    \rput[0.2](-1.5, 4.25){\psframebox*{\footnotesize $\hat{\xxx} = [s_4,s_2,s_1]$}}
	\psline{->}(-1.5, 4.8)(-1, 5.4)
    \psline{->}(-1.5, 4.8)(0.2, 5.2)
    \rput[0.2](-0.4, 3.35){\psframebox*{\footnotesize $\xxx = [s_1,s_4,s_3,s_2]$}}
    \rput[0.2](-0.4, 2.95){\psframebox*{\footnotesize $\hat{\xxx} = [s_4,s_3,s_2,s_1]$}}
    \psline{->}(0, 3.5)(1, 4.2)
    \psline{->}(0, 3.5)(2.25, 3.9)
    \psline{<->}(2.9, 1.72)(3.85, 1.72)
    \rput(3.3,1.72){\rnode{A}{}}
	\rput(1,2){\rnode{B}{\psframebox*{\footnotesize $\approx 1.25$ dB }}}
	\nccurve[angleB=0]{A}{B}
	\psellipse(1.17,5.35)(0.15,0.35)
	\rput[0.2](2.3, 5.95){\psframebox*{\footnotesize $\xxx = [s_3, s_3]$}}
    \rput[0.2](2.3, 5.55){\psframebox*{\footnotesize $\hat{\xxx} = [s_1, s_1]$}}

    \end{pspicture}
    \caption{The PEP for three different pairs of codewords $\xxx$ and $\hat{\xxx}$. Solid and dashed lines represent analytical PEP in~\eqref{eq:PEP} for the S-DEC and the B-DEC, resp. Filled and empty markers show simulation results for the S-DEC and the B-DEC, resp. The dotted line shows the exact PEP for the B-DEC (see~\secref{sec:ZcMod}).}
    \label{fig:loss125dB}
\end{figure}

\figref{fig:loss125dB} shows the analytical and the simulated PEP for the S-DEC and the B-DEC as functions of $d/\sigma_z$ for three different pairs of codewords $\xxx$ and $\hat{\xxx}$. We note that $d^2/\sigma^2_z$ is proportional to the SNR. Solid and dashed lines represent analytical PEP in~\eqref{eq:PEP} for the S-DEC and the B-DEC, respectively. For the codewords $\xxx = [s_3, s_3]$ and $\hat{\xxx} = [s_1, s_1]$ (circles), the dashed line coincides with the solid line. Filled markers represent simulation results for the S-DEC and are exactly on top of the corresponding solid lines. Empty markers show simulation results for the B-DEC. Empty squares and diamond agree well with the analytically predicted PEP; however, empty circles deviate significantly from the analytical prediction (which is based on the ZcMod approximation). We note that instead, empty circles agree well with the dotted line, which is briefly discussed in the next section.

\subsection{Zero-Crossing Approximation}\label{sec:ZcMod}

The exact PDFs of the L-values are superpositions of piece-wise Gaussian functions~\cite{Alvarado07d}. The ZcMod approximation uses only one Gaussian function to approximate the exact PDF. Although the ZcMod approximation has been shown to be good in terms of coded bit-error rate (BER) and mutual information~\cite[Sec.~5]{Benjillali07}, \cite[Sec.~III-C]{Alvarado07d}, a rigorous proof of its tightness is still missing. This is mainly because it requires to consider any pair of codewords. In the following, we show that the approximation is asymptotically tight for the codewords $\xxx = [s_3, s_3]$ and $\hat{\xxx} = [s_1, s_1]$ (circles in~\figref{fig:loss125dB}).

For the codewords $\xxx = [s_3, s_3]$ and $\hat{\xxx} = [s_1, s_1]$, $\Delta^{\setB}(\xxx, \hat{\xxx})$ in~\eqref{eq:dist_bdec} is a sum of two SMDs. When calculating the PEP, a convolution of the PDFs of these SMDs needs to be calculated.  The peculiarity of these SMDs, when $s_3$ is transmitted and $s_1$ is a competitor, is that their PDFs contain a Dirac delta function, which comes from the horizontal piece of the L-value function (see e.g., the  solid line in~\cite[Fig.~3b]{Alvarado10d}). When two such PDFs are convolved, the resulting PEP is not well approximated by the ZcMod approximation. The exact PEP calculated numerically using the exact PDF is shown with a dotted line in~\figref{fig:loss125dB} and, as expected, it coincides with the simulations for the B-DEC (empty circles).

\begin{figure}
    \newcommand{\scale}{0.85}
    \psfrag{xlabel}[cc][cB][\scale]{$d/\sigma_z$~[dB]}
    \psfrag{ylabel}[cc][ct][\scale]{PEP ratio}
    \centering
    \includegraphics[width=\columnwidth]{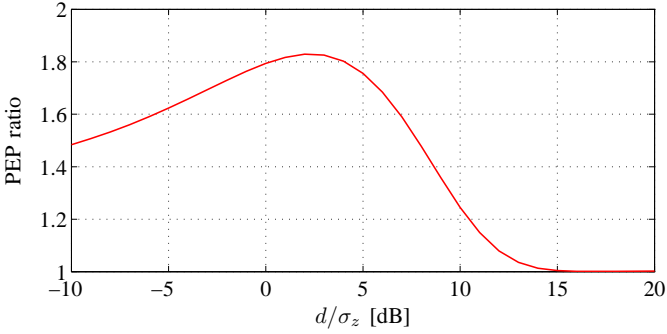}
    \caption{The ratio between the exact PEP (obtained numerically) using the exact PDF of the SMDs and the PEP predicted by the ZcMod approximation for the codewords $\xxx = [s_3, s_3]$ and $\hat{\xxx} = [s_1, s_1]$ (circles in~\figref{fig:loss125dB}).}
    \label{fig:pep_ratio}
\end{figure}

To study the asymptotic tightness of the ZcMod approximation, we show in \figref{fig:pep_ratio} the ratio between the exact PEP and the approximated PEP. This figure shows that for moderate SNR, the ZcMod approximation underestimates the PEP. However, the approximation is tight when $d/\sigma_{z} \rightarrow \infty$. This result was also verified analytically by considering upper and lower bounds on the exact PEP. Analogous results were obtained for other pairs of codewords. A behavior similar to what is shown in~\figref{fig:pep_ratio} will be  observed later on in~\secref{sec:simulations}.

\subsection{Asymptotic Pairwise Loss}\label{sec:asympt_comp}

Using \eqref{eq:PEP} and  \eqsref{A.PEP}{A.PEP2}, we define the asymptotic loss (when $d/\sigma_z\rightarrow \infty$) caused by the B-DEC (compared to the S-DEC) for any two pairs of codewords $\xxx$ and $\hat{\xxx}$ as
\begin{align}
\Loss(\xxx, \hat{\xxx}) 		& \triangleq 20\log_{10}\left(\frac{a^\setX(\xxx, \hat{\xxx})}{a^\setB(\xxx, \hat{\xxx})}\right).
\label{Loss}
\end{align}

The following theorem gives a bound on~\eqref{Loss}.

\begin{theorem}\label{Bound.Theorem}
For $4$-PAM with any Gray labeling, $\Loss(\xxx, \hat{\xxx})\leq1.25~\textrm{dB}$ for any two codewords $\xxx$ and $\hat{\xxx}$.
\end{theorem}

\begin{IEEEproof} Substituting the values in \tabsref{tab:S_distributions}{tab:B_distributions} into~\eqsref{A.PEP}{A.PEP2}, the normalized distances can be expressed as
\begin{align}\label{eq:A_SDEC}
a^\S(\xxx, \hat{\xxx}) &= \sqrt{\beta + 8w_{c}},\\
a^\B(\xxx, \hat{\xxx}) &= \beta^{-1/2}(\beta+ 2w_{c}),\label{eq:A_BDEC}
\end{align}
where
\begin{equation}\label{eq:beta}
	\beta = \sum_{\eee}{w_{\eee}\mu_{\eee}} = \sum_{\eee}{w_{\eee}\sigma^2_{\eee}}.
\end{equation}
The loss in \eqref{Loss} is then given by
\begin{equation}
\Loss(\xxx, \hat{\xxx}) =  20\log_{10}\left(\frac{\sqrt{\beta(\beta + 8w_{c})}}{\beta + 2w_{c}}\right).
    \label{eq:gain}
\end{equation}
The argument of the logarithm in~\eqref{eq:gain} is a positive function of $\beta$ and $w_{c}$ with a single maximum at $\beta = 4 w_{c}$. The maximum value is $\frac{2}{\sqrt{3}}$, which gives $\Loss(\xxx, \hat{\xxx}) \le 1.25$ dB.
\end{IEEEproof}

From the proof of~\theref{Bound.Theorem} it follows that the loss is zero if $w_c =0$ and it achieves its maximum if $\beta$ in~\eqref{eq:beta} is equal to $4w_{c}$. Using~\tabsref{tab:S_distributions}{tab:B_distributions}, it is easy to show that the latter condition is fulfilled for the pair of codewords $\xxx = [s_1, s_4, s_3, s_2]$ and $\hat{\xxx} = [s_4, s_3, s_2, s_1]$, and the asymptotic loss is $1.25$~dB, as illustrated by the simulation and analytical results (squares) in~\figref{fig:loss125dB}.

\section{Asymptotic Loss for Codes}

When all the codewords of a code are considered (e.g., in a union bound-type of expression~\cite[Ch.~4]{Bigl_Book}), only the pairs of codewords at minimum distance will define the high-SNR performance. The asymptotic loss for a given code $\setB$ can then be expressed as
\begin{equation}\label{eq:Loss4Code}
    \CLoss(\setB) \triangleq 20\log_{10}\left(\frac{\min_{\xxx \neq \hat{\xxx} \in \setX} a^\setX(\xxx, \hat{\xxx})}{\min_{\xxx \neq \hat{\xxx} \in \setX} a^\setB(\xxx, \hat{\xxx})}\right).
\end{equation}

In this section, we study the asymptotic loss in~\eqref{eq:Loss4Code}. We first consider an arbitrary linear code and then discuss a particular case of rate-$1/2$ CCs.

\subsection{Any Linear Code}
	The next corollary is a straightforward implication of~\theref{Bound.Theorem}.

\begin{corollary}
For $4$-PAM with any Gray labeling and any linear code, $\CLoss(\setB) \le 1.25$~dB. There exist CM codes for which this bound is exact.
\end{corollary}

\begin{IEEEproof}
The proof of the first statement follows directly from Theorem~\ref{Bound.Theorem} and \eqref{eq:Loss4Code}. To prove the second part, we give an example of such a code. Consider a linear code consisting of two codewords $\bbb_1 = [0,\, 0,\, 0,\, 0,\, 0,\, 0,\, 0,\, 0]$ and $\bbb_2 = [1,\, 0,\, 0,\, 1,\, 0,\, 1,\, 1,\, 1]$ used with $4$-PAM and \GC1. This corresponds to a CM code with two codewords $\xxx_1 = [s_1,\, s_1,\, s_1,\, s_1]$ and $\xxx_2 = [s_4,\, s_2,\, s_2,\, s_3]$. From~\tabsref{tab:S_distributions}{tab:B_distributions}, it follows that for these two codewords $\beta = 4 w_{c}$. Hence, $\CLoss(\setB) = \Loss(\xxx_1, \xxx_2) = 1.25$ dB.
\end{IEEEproof}

Even though linear codes with nonzero asymptotic loss exist, they are not very common due to their special structure, i.e., the closest paths should consist of a special combination of symbols. In what follows, we show that for some labelings and a wide range of linear codes, $w_{c} = 0$ for the codewords at minimum distance, and therefore, the asymptotic loss in~\eqref{eq:Loss4Code} is zero.

\begin{theorem}\label{the:GC3}
For $4$-PAM with \GC3 or \GC4 and any linear code, the loss $\CLoss(\setB) = 0$.
\end{theorem}

\begin{IEEEproof}
Consider the \GC3 labeling. Let $\xxx$ and $\hat{\xxx}$ be two different codewords of the code $\setX$ with corresponding binary codewords $\bbb, \hat{\bbb} \in \setB$, such that $w_{c}(\xxx, \hat{\xxx}) \neq 0$. For any linear code, $\bbb' = \bbb \oplus \bbb = [0,\dots,0]$ and $\hat{\bbb}' = \hat{\bbb} \oplus \bbb$ are also codewords of $\setB$ with corresponding $\xxx', \hat{\xxx}' \in \setX$. As $\bbb' \oplus \hat{\bbb}' = \bbb \oplus \hat{\bbb}$, we conclude that $w_{\eee}(\xxx', \hat{\xxx}') = w_{\eee}(\xxx, \hat{\xxx}),\, \forall \eee \in \setE$. From \tabsref{tab:S_distributions}{tab:B_distributions}, it is clear that $w_{c}(\xxx', \hat{\xxx}') = 0$, as $\xxx' = [s_2,s_2\dots,s_2]$. Using~\eqref{eq:A_SDEC} and the assumption that $w_{c}(\xxx, \hat{\xxx}) \neq 0$, we conclude that for the S-DEC
\begin{equation*}
a^{\setX}(\xxx, \hat{\xxx}) = \sqrt{\beta + 8w_{c}} > \sqrt{\beta} = a^{\setX}(\xxx', \hat{\xxx}').
\end{equation*}
Using~\eqref{eq:A_BDEC} we show, in a similar way, that for the B-DEC
\begin{equation*}
a^{\setB}(\xxx, \hat{\xxx}) = \beta^{-1/2}(\beta + 2w_c) > \sqrt{\beta} = a^{\setB}(\xxx', \hat{\xxx}').
\end{equation*}

We showed that $a(\xxx', \hat{\xxx}') < a(\xxx, \hat{\xxx})$ for both the S-DEC and the B-DEC. Hence, for any two codewords $\xxx$ and $\hat{\xxx}$ with $w_{c}(\xxx, \hat{\xxx}) \neq 0$, there always exist two other codewords $\xxx'$ and $\hat{\xxx}'$ with $w_{c}(\xxx', \hat{\xxx}') = 0$ at a smaller distance. The latter means that $w_{c}(\xxx, \hat{\xxx}) = 0$ for any pair of codewords $\xxx$ and $\hat{\xxx}$ at minimum distance, and hence, the loss in~\eqref{eq:Loss4Code} is zero. Similar reasoning directly applies to \GC4. This completes the proof.
\end{IEEEproof}

The peculiar property of \GC3 and \GC4 is that the all-zero label is assigned to one of the innermost constellation points, which guarantees that $\xxx = [s_2,s_2,\dots,s_2] \in \setX$. This is not the case for the \GC1 and \GC2 labelings, where the all-zero label is assigned to one of the outermost symbols. However, for these labelings it is still possible to define a family of codes for which the loss is also zero. This is done in the following theorem.

\begin{theorem}\label{the:BRGC_code}
For $4$-PAM with \GC1, the loss $\CLoss(\setB) = 0$ if the linear code $\setB$ contains a codeword $\bbb'' =[\bbb''[1], \dots, \bbb''[N]] \in \setB$, such that $b''_2[k] = 1,\, \forall k$. Similarly, for $4$-PAM with \GC2, $\CLoss(\setB) = 0$ if  $\bbb'' \in \setB$ and $b''_1[k] = 1,\,\forall k$.
\end{theorem}
\begin{IEEEproof}
First, we assume that \GC1 is used and a codeword $\bbb''$, such that $b''_2[k] = 1,\, \forall k$, belongs to the code $\setB$. Let $\xxx$ and $\hat{\xxx}$ be codewords of the code $\setX$ with corresponding binary codewords $\bbb, \hat{\bbb} \in \setB$, such that $w_{c}(\xxx, \hat{\xxx}) \neq 0$. For a linear code, $\bbb' = \bbb \oplus \bbb \oplus \bbb''$ and $\hat{\bbb}' = \hat{\bbb} \oplus \bbb \oplus \bbb''$ are also codewords of $\setB$ with corresponding $\xxx', \hat{\xxx}' \in \setX$. From \tabsref{tab:S_distributions}{tab:B_distributions}, it is clear that $w_{c}(\xxx', \hat{\xxx}') = 0$, as $\xxx' = [x'[1],\dots, x'[N]]$, where $x'[k] \in \{s_2, s_3\},\, \forall k$. The rest of the proof is similar to the proof of~\theref{the:GC3}. Swapping the first and the second bit positions in \GC1, we can analogously prove the second statement for \GC2. 
\end{IEEEproof}


\subsection{Rate-1/2 Convolutional Codes}
Bringing together the results for different labelings (\thesref{the:GC3}{the:BRGC_code}), the conclusion is as follows.
 
\begin{corollary}\label{cor:condition}
For $4$-PAM with any Gray labeling, $\CLoss(\setB) = 0$ if the linear code $\setB$ contains codewords $\bbb'', \bbb''' \in \setB$, such that $b''_1[k] = 1,\, \forall k$ and $b'''_2[k] = 1,\, \forall\, k$. 
\end{corollary}

Many codes satisfy the conditions in~Corollary~\ref{cor:condition}, for instance, all extended Hamming codes, all Reed-Muller codes, all extended BCH codes, and all extended Golay codes. All these codes include the all-one codeword. The codes are extended as they should be of an even length to match the constellation. For such codes, all the four Gray labelings are equivalent, in the sense that for a given binary code they produce four \emph{different} CM codes, with the \emph{same} minimum distance for both the S-DEC and the B-DEC.

Rate-$1/2$ CCs are of particular interest, as they allow an easy implementation of the ML decoder based on the Viterbi algorithm. In the following theorem, we show that all rate-$1/2$ CCs also give a zero asymptotic loss.

\begin{theorem}\label{theor.CC}
    For $4$-PAM with any Gray labeling and any rate-$1/2$ CC, $\CLoss(\setB) = 0$.
\end{theorem}

\begin{IEEEproof}
Any rate-$1/2$ CC $\setB$ can be generated by a generator matrix $\boldsymbol{G}(D) = [g_1(D),\,\,g_2(D)]$ \cite[Ch.~4.2]{Ryan_Lin_Book}, where $g_1(D)$ and $g_2(D)$ are nonzero generator polynomials over the binary field\footnote{We assume that any CC is realizable (see~\cite[Ch.~4.2]{Ryan_Lin_Book}) and such that $g_i(D) \neq 0$ for $i = 1,2$.}.  We assume that $g_1(D)$ defines odd bits of codewords $b_1[k]$, and $g_2(D)$ defines even bits $b_2[k]$. Any  generator matrix $\boldsymbol{G}(D)$ can be put in a systematic form $\boldsymbol{G}_{\mathrm{sys}}(D) = [1,\,\,g_2(D)/g_1(D)]$. Thus, an all-one input will produce a codeword where every odd bit is one, i.e., $\bbb''$, such that $b''_1[k] = 1,\, \forall k$. Analogously, any generator matrix $\boldsymbol{G}(D)$ can be put in the form $\boldsymbol{G}'_{\mathrm{sys}}(D) = [g_1(D)/g_2(D),\,\,1]$, which means that an all-one input produces a codeword where every even bit is one, i.e., $\bbb'''$, such that $b'''_2[k] = 1,\, \forall k$. The three generator matrices $\boldsymbol{G}(D)$, $\boldsymbol{G}_{\mathrm{sys}}(D)$, and $\boldsymbol{G}'_{\mathrm{sys}}(D)$ generate the same code, i.e., any rate-$1/2$ CC $\setB$ satisfies the conditions of Corollary~\ref{cor:condition}. This completes the proof.
\end{IEEEproof}

\begin{remark} \label{rem:8terms}
    Using a similar argument to the proof of~\theref{the:GC3}, we can show that for codes satisfying conditions in~Corollary~\ref{cor:condition}, $w_c(\xxx, \hat{\xxx}) = 0$ not only for codewords at minimum distance but also for the first eight terms in the distance spectrum. We therefore conclude that the bound developed in~\cite{Alvarado10d} is, in fact, a TCM union bound (at least for the first 8 terms) obtained from the spectrum of a \emph{binary} code.
\end{remark}

\subsection{Application: Optimal Bit-Wise Schemes}\label{sec:simulations}

In this section, we show how optimal bit-wise schemes can be found for rate-$1/2$ CCs. One approach is presented in~\cite{Alvarado10d}, where a search over all feedforward encoders was performed. The alternative approach we use here is to exploit the encoder equivalence shown in~\cite{Alvarado13}, which states that for CCs, different labelings can be grouped into classes that result in the same CM code $\setX$. In other words, the same CM code $\setX$ can be obtained by any labeling within a class used together with a properly modified convolutional encoder. This allows us to use the results reported in~\cite{Alvarado13} with the set-partitioning (SP) labeling~\cite{Unger82jan}.

For many constellations, including $4$-PAM, the SP and Gray labelings belong to the same class~\cite[Theorem~3]{Alvarado13}. Let $\setX$ be a CM code obtained by the CC with generator matrix $\boldsymbol{G}_{\mathrm{SP}}(D) = [g_1(D),\,\, g_2(D)]$ and $4$-PAM with the SP labeling given by $\qqq_{\mathrm{SP}} = [0,1,2,3]$. The same CM code $\setX$ can be obtained by $\boldsymbol{G}_{\mathrm{BRGC}}(D) = [g_1(D),\,\, g_1(D) + g_2(D)]$ and $4$-PAM with \GC1. We use this to obtain codes for the optimal bit-wise schemes, shown in~\tabref{tab:CC}, from codes for the optimal TCM schemes presented in~\cite[Table~III]{Alvarado13}. From now on, we use octal representation for the generator polynomials and omit the argument $D$ of the generator matrix. For memories $\nu = 2,3,4,6,7$, the codes in~\tabref{tab:CC} coincide with the codes in~\cite[Table~III]{Alvarado10d} ($\nu = 1, 8$ are not reported). For some $\nu$, there may be several encoders with identical performance, which explains the different codes for $\nu = 5$.

\begin{table}
    \centering
    \caption{Generator polynomials for rate-$1/2$ CCs that give optimal TCM encoders for $4$-PAM with the BRGC}
    \begin{tabular}{|c|c|c|c|}
        \hline
        $\nu$ & $\boldsymbol{G}$ & $\nu$ & $\boldsymbol{G}$\\
        \hline
        1& $[3,\, 2]$& 5& $[55,\, 51]$\\
        \hline
        2& $[7,\, 5]$ & 6& $[107,\, 135]$\\
        \hline
        3& $[13,\, 17]$ & 7& $[313,\, 235]$\\
        \hline
        4& $[23,\, 33]$ & 8& $[677,\, 515]$\\
        \hline
    \end{tabular}
    \label{tab:CC}
\end{table}

\begin{figure}
    \newcommand{\scale}{0.85}
    \psfrag{xlabel}[cc][cB][\scale]{${E_s}/{N_0}$~[dB]}%
    \psfrag{ylabel}[cc][ct][\scale]{BER}%
	\psfrag{legend111}[l][l][\scale]{ S-DEC}
	\psfrag{legend2}[l][l][\scale]{ B-DEC}%

    \centering
	\includegraphics[width=\columnwidth]{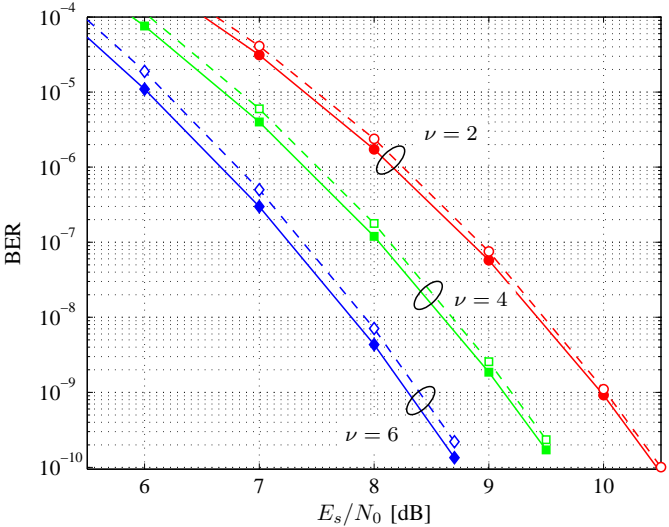}
	\vspace{-0.42cm}
	\begin{pspicture}(0,0)(0,0)
	\psset{linewidth=.02cm}
	\rput{135}(0.7,5.25){\psellipse(0,0)(0.12,0.25)}
	\rput(1.5, 5.6){\psframebox*{\footnotesize $\nu = 2$}}
	\rput{135}(1.2,3.45){\psellipse(0,0)(0.12,0.25)}
	\rput(1.9, 3.4){\psframebox*{\footnotesize $\nu = 4$}}
	\rput{135}(1.1,2.05){\psellipse(0,0)(0.12,0.25)}
	\rput(0.45, 1.6){\psframebox*{\footnotesize $\nu = 6$}}
	\end{pspicture}
    \caption{BER simulation results for rate-$1/2$ CCs in~\tabref{tab:CC} over the AWGN channel. the S-DEC and the B-DEC are shown with solid and dashed lines, respectively.}
    \label{fig:codes}
\end{figure}
\begin{figure}[t]
    \newcommand{\scale}{0.85}
    \psfrag{xlabel}[cc][cB][\scale]{${E_s}/{N_0}$~[dB]}%
    \psfrag{ylabel}[cc][ct][\scale]{BER ratio}%
    \psfrag{legend3}[c][c][\scale]{\hspace{-2mm}$\nu= 2$}
    \psfrag{legend2}[c][c][\scale]{\hspace{-2mm}$\nu= 4$}
    \psfrag{legend1}[c][c][\scale]{\hspace{-2mm}$\nu= 6$}
    \centering
	\includegraphics[width=\columnwidth]{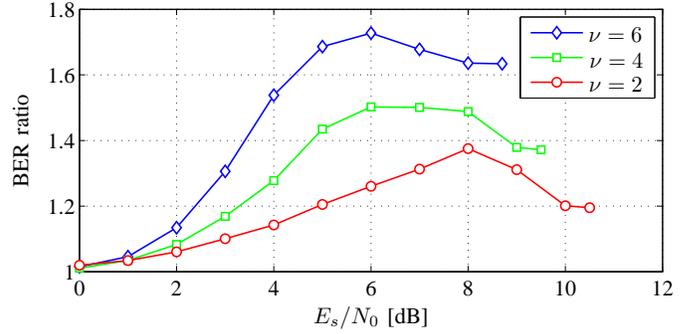}
    \caption{Ratios between the BER curves for the B-DEC and the S-DEC in~\figref{fig:codes}.}
    \label{fig:ratio_awgn}
\end{figure}

\figref{fig:codes} shows the S-DEC and the B-DEC performance for CCs with memories $\nu = 2,\,4,\,6$ in~\tabref{tab:CC}. As predicted by the results in~\secref{sec:asympt_comp}, the B-DEC gives rise to a higher probability of error at moderate SNRs (the loss is approximately $0.2$ dB). The gap between the B-DEC and the S-DEC decreases when the SNR increases, which is clearly seen from the curves marked with circles. As~\figref{fig:pep_ratio} suggests, the gap between the decoders is expected to be negligible at $d/\sigma_z \approx 15$ dB. This corresponds to $E_s/N_0 \approx 11$~dB, which is beyond our simulation capabilities. To support the fact that the gap does indeed disappear at high SNR, in \figref{fig:ratio_awgn} we show ratios between the BER curves. As we can see, the curves behave similarly to the curve in~\figref{fig:pep_ratio}, i.e., the curves converge to constants and high SNR, which confirms the asymptotic equivalence of the two decoders.

\section{Extensions}
\subsection{Flat Fading Channels}\label{sec:fading}

In this section, we discuss the performance of the S-DEC and the B-DEC over flat fading channels. The channel model in this case~is
\begin{equation}
    Y[k] = H[k] x[k] + Z[k],
\end{equation}
where $H[k]$ are channel coefficients, which are assumed to be known at the receiver.

For a given realization of the channel coefficients $\boldsymbol{h} = [h[1], h[2],\dots, h[N]]$, the ML decoding rule is given by~\eqref{eq:S-decoding_rule}, where $D^\S(\xxx)$ is now calculated as $D^\S(\xxx) = \sum_{k = 1}^{N}{\left(Y[k]-h[k]x[k]\right)^2}$. For the B-DEC, only the calculation of L-values changes compared to the Gaussian channel, i.e.,
\begin{multline}
L_j[k] = \frac{1}{2\sigma_z^2}\left[\min_{s \in \setS_{j, 0}}{(Y[k]-h[k]s)^2}\right. \\ -\left. \min_{s \in \setS_{j, 1}}{(Y[k]-h[k]s)^2}\right].
    \label{eq:max_log_rrl_fade}
\end{multline}

It can be easily shown that the PEP can be calculated as in~\eqref{eq:PEP_SDEC} or~\eqref{eq:PEP_BDEC}, where $\Delta(\xxx, \hat{\xxx})$ in this case is
\begin{align}
\Delta(\xxx, \hat{\xxx}) = \sum_{k = 1}^{N} h^2[k] \Lambda(x[k], \hat{x}[k]),
\end{align}
and $\Lambda(x[k], \hat{x}[k])$ are given by~\eqref{eq:cm_lambda} and~\eqref{eq:delta_llrs} for the S-DEC and the B-DEC, respectively. The PEP for given  $\xxx$ and $\hat{\xxx}$ now depends on the channel coefficients $\boldsymbol{h}$, i.e., it is given by
\begin{equation} \label{eq:PEP_fad}
    \PEP(\boldsymbol{h},\xxx, \hat{\xxx}) = \qfunc{a( \boldsymbol{h}, \xxx, \hat{\xxx})\frac{d}{\sigma_z}},
\end{equation}
where the normalized distance $a( \boldsymbol{h}, \xxx, \hat{\xxx})$ now incorporates the channel coefficients. Namely, the normalized distance for the S-DEC is given by
\begin{multline}\label{eq:dist_fad}
    a^{\S}(\boldsymbol{h}, \xxx, \hat{\xxx}) \\ = \frac{ \sum_{k \in \setK_{c}} h^2[k](\mu_{\S} - \mu_{[1,0]}) + \sum_{\eee \in \setE} \sum_{k \in \setK_{\eee}} h^2[k] \mu_{\eee}} {\sqrt{\sum_{k \in \setK_{c}} h^2[k](\sigma^2_{\S} - \sigma^2_{[1,0]}) + \sum_{\eee \in \setE} \sum_{k \in \setK_{\eee}} h^2[k] \sigma^2_{\eee}}},
\end{multline}
where 
\begin{equation*}
\setK_{\eee} = \left\{k\in \{1,...,N\}:\Phi^{-1}_{\mathcal{S}}(x[k]) \oplus \Phi^{-1}_{\mathcal{S}}(\hat{x}[k]) = \eee\right\}
\end{equation*}
and $\setK_{c}$ is the set of indices of pairs $(x[k], \hat{x}[k])$ in $\xxx$ and $\hat{\xxx}$ such that $(x[k], \hat{x}[k]) = (s_1, s_4)$ or $(x[k], \hat{x}[k]) = (s_4, s_1)$. We note that $|\setK_{\eee}| = w_{\eee}, \,\, \forall \eee \in \setE$ and $|\setK_{c}| = w_{c}$. The normalized distance $a^\B(\boldsymbol{h}, \xxx, \hat{\xxx})$ for the B-DEC can be obtained from~\eqref{eq:dist_fad} by replacing $\mu_\S$ and $\sigma_\S$ with $\mu_\B$ and $\sigma_\B$, respectively. The asymptotic loss can therefore be expressed as in~\eqref{Loss} using the distances defined above. This allows us to formulate the following theorem. 

\begin{theorem}\label{Bound.Theorem_fad}
For $4$-PAM and any Gray labeling, $\Loss(\xxx, \hat{\xxx})\leq1.25~\textrm{dB}$ for any two codewords $\xxx$ and $\hat{\xxx}$ and any given channel realization $\boldsymbol{h}$.
\end{theorem}

\begin{IEEEproof}
For a given channel realization $\boldsymbol{h}$, the asymptotic loss can be expressed similarly to~\eqref{eq:gain} as
\begin{equation}\label{eq:loss_fad}
    \Loss(\xxx, \hat{\xxx}) =  20\log_{10}\left(\frac{\sqrt{\beta(\beta + 8\alpha)}}{\beta + 2\alpha}\right),
\end{equation}
where
\begin{align*}
	\alpha &= \sum_{k \in \setK_{c}} h^2[k],\\
	\beta &= \sum_{\eee \in \setE} \sum_{k \in \setK_{\eee}} h^2[k] \mu_{\eee}.
\end{align*}
Analogously to the proof of~\theref{Bound.Theorem}, we can show that the maximum value of \eqref{eq:loss_fad} is 1.25 dB when $\beta = 4\alpha$.
\end{IEEEproof}

An adequate performance measure for fading channels is the average PEP, where the average is taken over the fading distribution. Formally, the average PEP is defined as $\overline{\PEP}(\xxx, \hat{\xxx}) = \mathsf{E}_{\HHH}\left\{\PEP(\HHH, \xxx, \hat{\xxx}) \right\}$, 
where $\mathsf{E}_{\HHH}\{ \cdot \}$ denotes expectation over $\HHH$. The next corollary gives a result for the average PEP and follows directly from~\theref{Bound.Theorem_fad}.
\begin{corollary}
For 4-PAM and any Gray labeling, the average asymptotic loss for the two decoders is $\leq 1.25$ dB for any two codewords $\xxx$ and $\hat{\xxx}$.
\end{corollary}

More precise conclusions about the average PEP can be drawn if the distribution of $\HHH$ is specified. However, we note that if $w_c(\xxx, \hat{\xxx}) = 0$, the two decoders give the same $\PEP(\boldsymbol{h}, \xxx, \hat{\xxx})$, and hence, the same $\overline{\PEP}(\xxx, \hat{\xxx})$, regardless of the distribution of $\HHH$.

\begin{figure}
    \newcommand{\scale}{0.85}
    \psfrag{xlabel}[cc][cB][\scale]{${E_s}/{N_0}$~[dB]}%
    \psfrag{ylabel}[cc][ct][\scale]{BER}%
	\psfrag{legend111}[l][l][\scale]{ S-DEC}
	\psfrag{legend2}[l][l][\scale]{ B-DEC}%
    \centering
    \includegraphics[width=0.993\columnwidth]{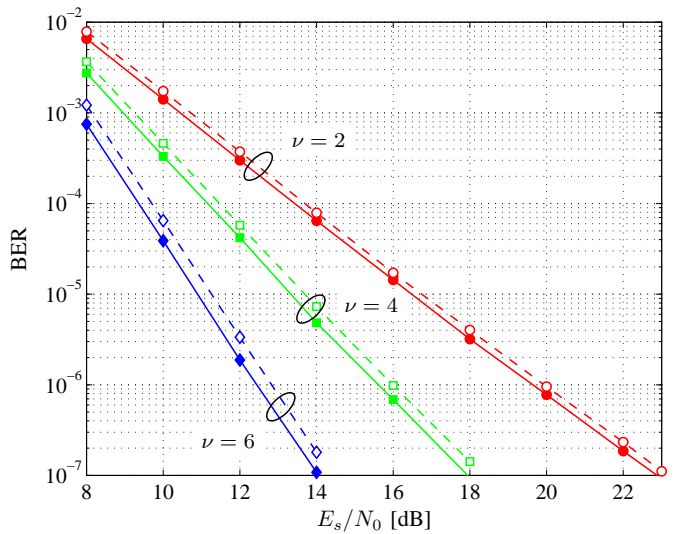}
    	\vspace{-0.42cm}
	\begin{pspicture}(0,0)(0,0)
	\psset{linewidth=.02cm}
	\rput{135}(-1.1,5.25){\psellipse(0,0)(0.12,0.25)}
	\rput(-0.3, 5.6){\psframebox*{\footnotesize $\nu = 2$}}
	\rput{135}(-0.4,3.35){\psellipse(0,0)(0.12,0.25)}
	\rput(0.4, 3.4){\psframebox*{\footnotesize $\nu = 4$}}
	\rput{135}(-0.8,2.05){\psellipse(0,0)(0.12,0.25)}
	\rput(-1.5, 1.6){\psframebox*{\footnotesize $\nu = 6$}}
	\end{pspicture}
    \caption{BER simulation results for rate-$1/2$ CCs in~\tabref{tab:CC} over the i.i.d. Rayleigh fading channel. the S-DEC and the B-DEC are shown with solid and dashed lines, resp.}
    \label{fig:codes_ray}
\end{figure}

The performance analysis for codes over the AWGN channel in~\thesref{the:GC3}{the:BRGC_code} showed that if $\xxx, \hat{\xxx} \in \setX$ are such that $w_c(\xxx, \hat{\xxx}) \neq 0$, in many cases we can find two other codewords $\xxx', \hat{\xxx}' \in \setX$, such that $w_c(\xxx', \hat{\xxx}') = 0$ and $w_{\eee}(\xxx, \hat{\xxx}) = w_{\eee}(\xxx', \hat{\xxx}')$ for $\eee \in \setE$. The codewords $\xxx$ and $\hat{\xxx}$ have therefore negligible impact on the code performance over the AWGN channel due to a larger than minimum distance between them. Even though the distance may not be the main parameter determining the average PEP for a flat fading channel, one could argue that the codewords $\xxx$ and $\hat{\xxx}$ are less relevant for the code performance than the codewords $\xxx'$ and $\hat{\xxx}'$. Bearing this in mind, we conjecture that, for linear codes, the B-DEC and the S-DEC should perform very similarly over flat fading channels, regardless of the distribution of $\HHH$. This conjecture is supported by the simulation results shown in~\figref{fig:codes_ray} presenting the BER of the S-DEC and the B-DEC over the independent identically distributed (i.i.d.) Rayleigh fading channel for CCs with $\nu = 2,\,4,\,6$ from~\tabref{tab:CC}. An important parameter for the average PEP for this channel is the number of different symbols between the two codewords~\cite[Sec.~III]{Divsalar88}, \cite[Sec.~I]{KCavers92}, which can be calculated as $\sum_{\eee \in \setE}w_{\eee}(\xxx, \hat{\xxx})$. Hence, codewords with $w_c(\xxx, \hat{\xxx}) \neq 0$ may visibly contribute to the performance. This explains a difference between the S-DEC and the B-DEC in~\figref{fig:codes_ray} even at high SNR.

\begin{remark}
	For the i.i.d. Rayleigh fading channel, an interleaver could be added between the MOD and the B-ENC in~\figref{fig:block_scheme} in order to increase the number of different symbols between the codewords, and hence, improve the performance. Although ML decoding is theoretically still possible in this case, it is too complex to implement, and thus, the B-DEC is preferred in practice. Even though we cannot obtain simulation results for the ML decoder, we conjecture that its performance is very similar to that of the B-DEC.
\end{remark}

\subsection{$64$-QAM Constellation}

The results presented in the paper can be used to predict the performance of some popular CM schemes that use other constellations than $16$-QAM. To illustrate this, we chose a CM scheme with $64$-QAM formed as the direct product of two $8$-PAM constellations. For each of the $8$-PAM constellations, we use the coding scheme devised by Ungerboeck where an \emph{uncoded} bit is assigned to the most protected bit position in the labeling. Below we show that this coding scheme can be seen as a coding scheme with $4$-PAM.

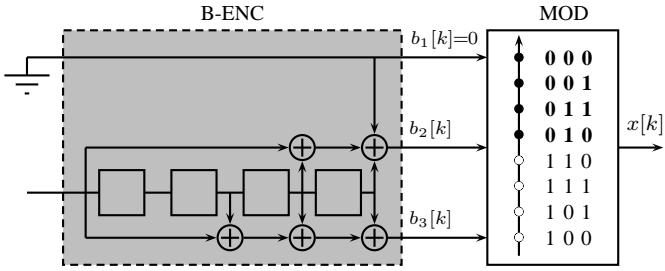
\begin{figure}[tp]
	\centering
	\psset{unit=.6cm}
    \footnotesize
	\begin{pspicture}(2.3,1)(16.9,7.3)
		\pspolygon[linearc=0, fillstyle = solid, fillcolor = lightgray, linestyle=dashed,dash=3pt 2pt](3.6, 1.4)(3.6, 6.6)(11.1, 6.6)(11.1, 1.4)
        \pspolygon[linearc=0, fillstyle = solid, fillcolor = white](13, 1.4)(13, 6.6)(15.9, 6.6)(15.9, 1.4)
		\pspolygon[linearc=0, fillstyle = solid, fillcolor = lightgray](4.4, 2.5)(4.4, 3.5)(5.4, 3.5)(5.4, 2.5)
		\pspolygon[linearc=0, fillstyle = solid, fillcolor = lightgray](6, 2.5)(6, 3.5)(7, 3.5)(7, 2.5)
		\pspolygon[linearc=0, fillstyle = solid, fillcolor = lightgray](7.6, 2.5)(7.6, 3.5)(8.6, 3.5)(8.6, 2.5)
		\pspolygon[linearc=0, fillstyle = solid, fillcolor = lightgray](9.2, 2.5)(9.2, 3.5)(10.2, 3.5)(10.2, 2.5)
		\psline{->}(13.7, 1.6)(13.7, 6.5)	
		\psdot[dotstyle = o](13.7, 2.01)
		\psdot[dotstyle = o](13.7, 2.58)
		\psdot[dotstyle = o](13.7, 3.15)
		\psdot[dotstyle = o](13.7, 3.72)
		\psdot(13.7, 4.29)
		\psdot(13.7, 4.86)
		\psdot(13.7, 5.43)
		\psdot(13.7, 6)

		\rput(14.8, 2.01){1 0 0}
		\rput(14.8, 2.58){1 0 1}
		\rput(14.8, 3.15){1 1 1}
		\rput(14.8, 3.72){1 1 0}
		\rput(14.8, 4.29){\bf 0 1 0}
		\rput(14.8, 4.86){\bf 0 1 1}
		\rput(14.8, 5.43){\bf 0 0 1}
		\rput(14.8, 6){\bf 0 0 0}
		
		\psline{-}(2.8, 6)(2.8, 5.6)
		\psline{-}(2.3, 5.6)(3.3, 5.6)
		\psline{-}(2.55, 5.4)(3.05, 5.4)
		\psline{-}(2.75, 5.2)(2.85, 5.2)
		\psline{->}(2.8, 6)(13, 6)        
		\psline{-}(2.8, 3)(4.4, 3)
		\psline{-}(5.4, 3)(6, 3)
		\psline{-}(7, 3)(7.6, 3)
		\psline{-}(8.6, 3)(9.2, 3)
		\psline{-}(10.2, 3)(10.5, 3)
		\pscircle(10.5,4){0.3}
		\psline{-}(10.5, 3.8)(10.5, 4.2)				
		\psline{-}(10.3, 4)(10.7, 4)
		\pscircle(10.5,2){0.3}
		\psline{-}(10.5, 1.8)(10.5, 2.2)				
		\psline{-}(10.3, 2)(10.7, 2)
		
		\pscircle(8.9,4){0.3}
		\psline{-}(8.9, 3.8)(8.9, 4.2)				
		\psline{-}(8.7, 4)(9.1, 4)
		\pscircle(8.9,2){0.3}
		\psline{-}(8.9, 1.8)(8.9, 2.2)				
		\psline{-}(8.7, 2)(9.1, 2)
		\pscircle(7.3,2){0.3}
		\psline{-}(7.3, 1.8)(7.3, 2.2)				
		\psline{-}(7.1, 2)(7.5, 2)
		\psline{-}(4.1, 3)(4.1, 4)
		\psline{-}(4.1, 3)(4.1, 2)
		\psline{->}(4.1, 4)(8.6, 4)
		\psline{->}(4.1, 2)(7, 2)
		\psline{->}(7.6, 2)(8.6, 2)
		\psline{->}(9.2, 2)(10.2, 2)
		\psline{->}(9.2, 4)(10.2, 4)
		\psline{->}(10.8, 4)(13, 4)
		\psline{->}(10.8, 2)(13, 2)
		\psline{<-}(10.5, 4.3)(10.5, 6)
		\psline{<->}(10.5, 2.3)(10.5, 3.7)
		\psline{<->}(8.9, 2.3)(8.9, 3.7)
		\psline{<-}(7.3, 2.3)(7.3, 3)
		\psline{->}(15.9, 4)(16.9, 4)
		\rput(12.05, 6.4){\scriptsize $b_1[k]\!\!=\!\!0$}
		\rput(11.75, 4.4){\scriptsize $b_2[k]$}
		\rput(11.75, 2.4){\scriptsize $b_3[k]$}
		\rput(16.5, 4.5){$x[k]$}
		\rput(14.7, 7){MOD}
		\rput(7.35, 7){B-ENC}
	\end{pspicture}
\caption{A CM scheme with an 8-PAM constellation labeled by the BRGC and a convolutional encoder $\boldsymbol{G}_{\mathrm{BRGC}} = [1, 1, 0; 0, 23, 27]$. If $b_1 = 0$, then only half of constellation points (highlighted) will be used for transmission.}
\label{fig:PAM8}
\end{figure}

\begin{figure}
    \newcommand{\scale}{0.85}
    \psfrag{xlabel}[cc][cB][\scale]{$E_b/N_0$~[dB]}
    \psfrag{ylabel}[cc][ct][\scale]{BER}
    \psfrag{Legend1111111111111111}[l][l][\scale][\scale]{Rate-$2/3$ Ungerboeck}
	\psfrag{Legend2}[l][l][\scale][\scale]{Rate-$1/3$ CC}
    \centering
    \vspace{0.08cm}
    \includegraphics[width=\columnwidth]{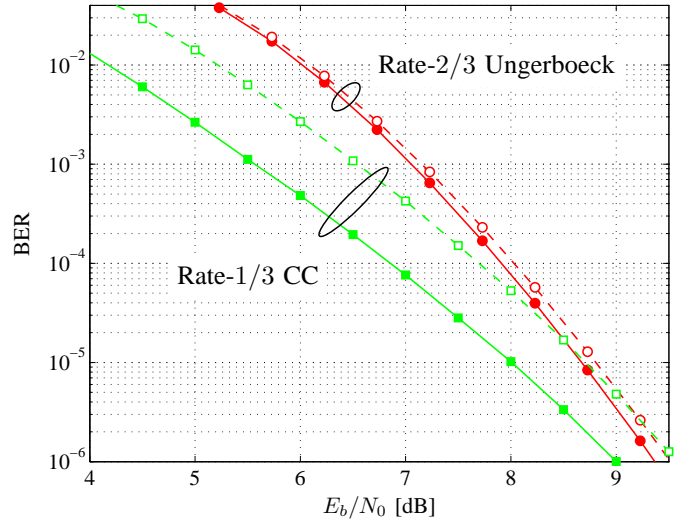}
	\vspace{-0.42cm}
	\begin{pspicture}(0,0)(0,0)
	\psset{linewidth=.02cm}
	\rput{135}(0,6){\psellipse(0,0)(0.12,0.25)}
	\rput(2, 6.4){\psframebox*{Rate-$2/3$ Ungerboeck}}
	\rput{135}(0.1,4.6){\psellipse(0,0)(0.12,0.65)}
	\rput(-1.3, 3.6){\psframebox*{Rate-$1/3$ CC}}
	\end{pspicture}
    \caption{BER simulation results for CM schemes with $8$-PAM over the AWGN channel. The S-DEC and the B-DEC are shown with solid and dashed lines, resp. Rate-2/3 Ungerboeck refers to the encoder in~\figref{fig:PAM8} and Rate-1/3 CC to the encoder with $\boldsymbol{G} = [25, 33, 37]$.}
    \label{fig:pam8}
\end{figure}

We chose a rate-$2/3$ CC with $\nu = 4$ and generator matrix $\boldsymbol{G}_{\mathrm{SP}} = [1, 0, 0; 0, 23, 4]$ (borrowing the notation from~\cite{Alvarado13}) with the SP labeling~\cite[Table~IV]{Alvarado13} to produce a CM code $\setX$. As discussed in \ref{sec:simulations}, the same code $\setX$ can be obtained by using the BRGC  together with the binary code $\setB$ generated by $\boldsymbol{G}_{\mathrm{BRGC}} = [1, 1, 0; 0, 23, 27]$. \figref{fig:PAM8} shows the described CM encoder. 
We consider a set of codewords $\setX_0 \subset \setX$, which can be produced by the CM encoder if all odd information bits are set to zero. The set of codewords $\setX_0$ can be seen as obtained by the concatenation of the code generated by $\boldsymbol{G}' = [23, 27]$ with a Gray-labeled $4$-PAM constellation, as highlighted in~\figref{fig:PAM8}. We can build tables similar to~\tabsref{tab:S_distributions}{tab:B_distributions} for $8$-PAM and show that $\setX_0$  captures most of distance properties of the original code $\setX$. We therefore expect the relative performance of the S-DEC and the B-DEC to be similar to that of $4$-PAM, i.e., we expect a small gap between the S-DEC and the B-DEC at moderate SNR. This is supported by the curves with circles in \figref{fig:pam8}, showing the BER performance for the described CM scheme over the AWGN channel. We conjecture that the decoders are asymptotically equivalent.

We cannot rely on the $4$-PAM analysis, however, when the most protected bit position is also encoded. As an example, we chose the best known binary rate-$1/3$ CC~\cite{Chang97, Bocharova97} for $\nu = 4$ with the generator matrix $\boldsymbol{G} = [25, 33, 37]$. Square markers in~\figref{fig:pam8} show the BER performance of the S-DEC and the B-DEC over the Gaussian channel for this coding scheme and demonstrate a significant difference between the two decoders.

\section{Conclusions} \label{sec:conclusions}

In this paper, we compared the ML symbol-wise decoder and a suboptimal bit-wise decoder based on max-log L-values. It was shown that asymptotically, the loss caused by the use of the suboptimal bit-wise decoder is bounded, and in many cases  equal to zero. The bit-wise decoder studied in this paper corresponds to the bit-interleaved coded modulation paradigm and is widely used in many wireless communication standards. The results in this paper can be seen as a theoretical justification for its use.


The analysis presented in this paper considered a $16$-QAM constellation labeled by any Gray labeling. Numerical results for $64$-QAM were also presented. These results support the  conjecture that the asymptotic equivalence between symbol-based and bit-based decoders may also be true in other cases. A rigorous analysis for other multilevel modulations is left for future investigation.

\bibliographystyle{IEEEtran}
\bibliography{MyBibliography}

\end{document}